\input harvmac
\def\np#1#2#3{Nucl. Phys. B {#1} (#2) #3}
\def\pl#1#2#3{Phys. Lett. B {#1} (#2) #3}
\def\plb#1#2#3{Phys. Lett. B {#1} (#2) #3}

\def\physrev#1#2#3{Phys. Rev. D {#1} (#2) #3}

\def\ev#1{\langle#1\rangle}
\def\V{{\cal V}}
\def\dim{{\rm dim}}
\def\f#1#2{\textstyle{#1\over #2}}
\def\mH{{\cal M}_H}
\def\mI{{\cal M}_{Inst}}
\def\FI{Fayet-Iliopoulos}
\def\cA{{\cal A}}
\def\G{{\cal H}}

\def\f#1#2{\textstyle{#1\over #2}}
\def\drawbox#1#2{\hrule height#2pt 
        \hbox{\vrule width#2pt height#1pt \kern#1pt 
              \vrule width#2pt}
              \hrule height#2pt}

\def\Fund#1#2{\vcenter{\vbox{\drawbox{#1}{#2}}}}
\def\Asym#1#2{\vcenter{\vbox{\drawbox{#1}{#2}
              \kern-#2pt       
              \drawbox{#1}{#2}}}}
 
\def\fund{\Fund{6.5}{0.4}}
\def\asym{\Asym{6.5}{0.4}}
\batchmode
  \font\bbbfont=msbm10
\errorstopmode
\newif\ifamsf\amsftrue
\ifx\bbbfont\nullfont
  \amsffalse
\fi
\ifamsf
\def\IR{\hbox{\bbbfont R}}
\def\IC{\hbox{\bbbfont C}}

\def\IZ{\hbox{\bbbfont Z}}
\def\IF{\hbox{\bbbfont F}}
\def\IP{\hbox{\bbbfont P}}
\else
\def\IR{\relax{\rm I\kern-.18em R}}
\def\IZ{\relax\ifmmode\hbox{Z\kern-.4em Z}\else{Z\kern-.4em Z}\fi}
\def\IF{\relax{\rm I\kern-.18em F}}
\def\IP{\relax{\rm I\kern-.18em P}}
\fi
\def\R{{\cal R}}
\def\P{{\cal P}}
\def\C{{\cal C}}
\lref\wsmall{E. Witten, 
hep-th/9511030, \np{460}{1995}{541.}}
\lref\dmw{M. Duff, R.  Minasian, and E. Witten, 
hep-th/9601036, \np{465}{1996}{413}.}
\lref\sw{N. Seiberg and E. Witten,  hep-th/9603003, \np{471}{1996}{121}.}
\lref\dlpt{M.J. Duff, H. Lu, and C.N. Pope, 
hep-th/9603037, \plb{378}{1996}{101}.}
\lref\gh{O. Ganor and A. Hanany,  hep-th/9602120.}
\lref\wsd{E. Witten,  hep-th/9609159, Mod. Phys. Lett. A11 (1996)
2649.}
\lref\Wcomm{E. Witten,  
hep-th/9507121, Proc. of Strings '95, editors I. Bars et. al., World
Scientific, 1996.}
\lref\AS{M.F. Atiyah and I.M. Singer, Ann. Math. 87 (1968) 485,
Ann. Math. 93 (1971) 119.} 
\lref\APS{M.F. Atiyah, V.K. Patodi, and I.M. Singer, 
Math. Proc. Camb. Phil. Soc. 77 (1975) 43; 77 (1975) 405; 79 (1976)
71.}
\lref\horwit{P. Horava and E. Witten, hep-th/9510209,
\np{460}{1996}{506}.} 
\lref\cjrm{C. V. Johnson and R. C. Myers, hep-th/9610140.}
\lref\slansky{R. Slansky, 
Physics Reports 79 (1981) 1.}
\lref\mdegs{M.R. Douglas,  hep-th/9612126.}
\lref\aspinB{P.S. Aspinwall,  hep-th/9507012, \plb{357}{1995}{329}.}
\lref\aspin{P. S. Aspinwall,  hep-th/9612108.}
\lref\Sanom{J. H. Schwarz,  hep-th/9512053, \pl{371}{1996}{223}.}
\lref\sw{N. Seiberg and E. Witten,  hep-th/9603003, \np{471}{1996}{121}.}

\lref\Betal{M. Berkooz, R.G. Leigh, J. Polchinski, J. Schwarz,
N. Seiberg, and E. Witten,  hep-th/9605184, \np{475}{1996}{115}.}
\lref\sdfp{N. Seiberg,  hep-th/9609161.}
\lref\obrane{K. Intriligator,  hep-th/9702038, Nucl Phys. B to
appear}
\lref\kron{P.B. Kronheimer, Jour. Differential Geometry, 
{\bf 29} (1989) 665.}
\lref\kn{P. B. Kronheimer and H. Nakajima, Math. Ann. 288 (1990) 263.}
\lref\wbound{E. Witten, 
hep-th/9510135, \np{460}{1996}{335}.}
\lref\GJ{E. G. Gimon and C. V. Johnson, 
hep-th/9604129, \np{477}{1996}{715}.}
\lref\mrdbranes{M. Douglas, hep-th/9512077.}
\lref\jbki{J. D. Blum, K. Intriligator,  hep-th/9605030.}

\lref\collins{M.J. Collins, {\it Representations characters of finite
groups,} Cambridge Studies in Advanced Mathematics 22, Cambridge
University Press, 1990.}
\lref\dm{M. Douglas and G. Moore,  hep-th/9603167.}
\lref\GP{E. G. Gimon and Polchinski,  hep-th/9601038,
\physrev{54}{1996}{1667}.}
\lref\Julie{J. D. Blum,  hep-th/9608053, \np{486}{1997}{34.}}
\lref\BluZaff{J. D. Blum and A. Zaffaroni,  \plb{387}{1996}{71}.}
\lref\swed{U.H. Danielsson, G. Feretti, J. Kalkkinen, and
P. Stjernberg, hep-th/9703098.}
\lref\bervaf{M. Bershadsky and C. Vafa, hep-th/9703167.}
\lref\sopen{A. Strominger, hep-th/9512059, \plb{383}{1996}{44}.}
\lref\witfb{E. Witten, hep-th/9512219, \np{463}{1996}{383}.}
\Title{hep-th/9705044, IASSNS-HEP-97/40}
{\vbox{\centerline{New Phases of String Theory and 6d RG Fixed Points}
\centerline{via Branes at Orbifold Singularities}}}
\medskip
\centerline{Julie D. Blum and 
Kenneth Intriligator\footnote{${}^*$}{On leave 1996-1997
{}from Department of Physics, University of California, San Diego.}}
\vglue .5cm
\centerline{School of Natural Sciences}
\centerline{Institute for Advanced Study}
\centerline{Princeton, NJ 08540, USA}

\bigskip
\noindent

We discuss type II and type I branes at general $ADE$ type orbifold
singularities.  We show that there are new phases of type I or
heterotic string theory in six dimensions, involving extra tensor
multiplets, which arise when small instantons sit on orbifold
singularities.  The theories with extra tensor multiplets are
explicitly constructed via orientifolds.  The world-volume theories in
type IIB or type I five-branes at orbifold singularities lead to the
existence of several infinite classes of six dimensional, interacting,
renormalization group fixed point theories.

\Date{5/97}                                   

\newsec{Introduction}

The small $E_8$ instanton is qualitatively different from the small
$SO(32)$ instanton.  In both cases, the small instanton is at a point
$p\in \IR ^4$ and thus yields a 5+1 dimensional world volume theory
with minimal ${\cal N}=(1,0)$ supersymmetry (8 supercharges).  In the
$SO(32)$ case, the 6d world-volume theory is a $Sp(1)$ gauge theory
\wsmall, which is IR free. On the other hand, the $E_8$ case is more
interesting: at the point in the moduli space where the $E_8$
instanton is of zero size, there is a transition to another ``Coulomb
branch'' moduli space of the theory, labeled by the expectation value
$\ev{\Phi }\in \IR ^+$ of a real scalar in a tensor multiplet.  The
branch with the tensor multiplet corresponds in $M$ theory to a phase
in which the zero size instanton has become a five-brane which has
moved a distance $\ev{\Phi }$ into the 11th dimension \refs{\dmw, \sw,
\dlpt}.  At the transition point between the two branches, where
$\ev{\Phi }\rightarrow 0$, there appears to be a ``tensionless
string''
\refs{\gh, \sw, \dlpt, \wsd} -- though there is evidence that it is an
interacting {\it local quantum field theory} at a non-trivial RG fixed
point.  The existence of other non-trivial fixed points of the
renormalization group in six dimensions was discussed in \refs{\sdfp,
\obrane, \swed,
\bervaf}.

A situation similar to that of the small $E_8$ instanton occurs when
the small $SO(32)$ instanton sits, not at a smooth point on the
transverse $\IR ^4$ but, rather, at a $\IZ _M$ orbifold singularity.
There are new phases with extra tensor multiplets and the world-volume
theory leads to non-trivial 6d super-conformal field theories \obrane.
For a $\IZ _2$ orbifold singularity, the new phase has a single extra
tensor multiplet and was found via $F$ theory in \aspin.  The new
phases for $\IZ _M$ singularities were given an orientifold
construction in \jbki.

Here we consider what happens when small $SO(N)$ instantons sit on
general orbifold singularities.  The result will be many more new
phases for the $SO(32)$ heterotic or type I string theories.  In
addition, the world-volume theories will lead to the existence of an
infinite number of new classes of non-trivial 6d superconformal field
theories.  We also point out that type IIB five-branes at orbifold
singularities lead to an infinite number of classes of non-trivial 6d
superconformal field theories.

In the next section we discuss some general preliminaries.  In the
first subsection, we review ALE spaces and their hyper-Kahler quotient
construction \kron.  As discussed in \refs{\dm, \cjrm}, these theories
have a physical realization via branes on the corresponding ALE space.
In the next subsection we discuss general aspects of instantons on ALE
spaces.  In sect. 3 we briefly review the situation for $U(N)$
instantons on ALE spaces and the hyper-Kahler quotient construction of
their moduli spaces \kn.  It is worth emphasizing that the
construction of \kn\ applies {\it only} for gauge group $U(N)$.  In
sect. 3.2 we mention the orbifold construction of these theories and
how tadpole conditions \jbki\ yield a relation of \kn.  Using a
relation \jbki\ between first Chern classes and the NSNS $B$ field,
this allows us to determine the value of the $B$ field in the orbifold
construction.

In sect. 4 we discuss $K$ IIB five-branes at a $\IC ^2/\Gamma _G$
orbifold singularity.  The 6d world-volume theory is a ${\cal
N}=(1,0)$ supersymmetric theory with gauge group and matter content
given by a ``moose\foot{in the terminology of
\ref\HGmoose{H. Georgi, \np{266}{1986}{274}.}.}''  (or ``quiver'') 
diagram, which is the extended Dynkin diagram of the $ADE$ group $G$
corresponding to the $\Gamma _G\subset SU(2)$, along with
$r=$rank$(G)$ extra tensor multiplets. The case $\Gamma _G=\IZ _2$
gives a theory mentioned in \Sanom\ as a solution of the anomaly
cancellation conditions.  We argue that these theories yield
six-dimensional, interacting, scale invariant theories for every
discrete $\Gamma _G\subset SU(2)$, and for any $K$.

In sect. 5 we discuss $SO(N)$ instantons on $\IC ^2/\Gamma _G$ ALE
spaces.  We present supersymmetric gauge theories whose Higgs branches
we argue give hyper-Kahler quotient constructions of the moduli space
of $SO(N)$ instantons on general ALE spaces, albeit with a caveat
about some of the ALE blowing up modes being frozen to zero.  The
theories are based on particular products of classical gauge groups
arranged according to moose diagrams which are again related to the
extended Dynkin diagrams of the $ADE$ gauge group $G$. This
construction applies for any $SO(N)$. These particular theories,
however, could not arise physically in the world-volume of D5 branes
at the orbifold singularity, as they suffer from deadly 6d gauge
anomalies.  For the special case of $SO(32)$ there is a modified
version of the gauge theories, with altered relations between the
gauge groups and matter content and also with additional tensor
multiplets, which we argue does arise physically in the world-volume of
the D5 branes.  These world-volume theories arise in a ``Coulomb
branch'' of the full moduli space, in which $29n_T$ hypermultiplets
are traded for $n_T$ tensor multiplets, much as in the small $E_8$
instanton.  The origin of the Coulomb branch yields, for every ALE
space, an infinite family of non-trivial, six dimensional,
interacting, scale invariant theories.

In sect. 6 we give an orientifold construction of these theories,
showing directly the existence of the new phases with extra tensor
multiplets.   

In sect. 7 we discuss other 6d gauge theories, which are similar to
those discussed in sect. 5, but without coupling between the
five-branes and the nine-branes.  The five-branes thus do not have an
interpretation as small instantons, and strings ending on these branes
do not have Dirichlet boundary conditions but rather twisted boundary
conditions.  Including these branes may resolve the puzzle of anomaly
cancellation in the theories discussed by \Julie.  One of these
theories, with gauge group given by the $A_1$ extended Dynkin diagram
moose, was previously found by \Sanom.  In sect. 7.1 we discuss the
orientifold construction of this theory.

Some general observations about anomaly restrictions on 6d gauge
theories based on product groups are made in sect. 8.  Finally, in an
appendix we include a few relevant facts about the representations of
the discrete subgroups of $SU(2)$.

{\bf Note added}: Related results concerning small $Spin (32)$
instantons at $D$ and $E$ type singularities have been independently
obtained via $F$ theory in \ref\aspmor{P. S. Aspinwall and
D. R. Morrison, RU-97-29, IASSNS-HEP-97/46, to appear.}.  We thank the
authors of \aspmor\ for comparing their results with ours, thereby
finding a typo in the original version of this paper concerning the
representations of $\Gamma _{E_7}$.

\newsec{General preliminaries}
\subsec{ALE spaces and branes}

The ALE spaces are solutions of Einstein's equations described as
possibly blown up orbifolds $\IC ^2/\Gamma _G$, with $\Gamma _G\subset
SU(2)$ acting on the $S^3$ at infinity.  The discrete $SU(2)$
subgroups: cyclic, dihedral, tetrahedral, octahedral, and icosahedral,
have a famous correspondence with, respectively, the $A_r$, $D_r$,
$E_6$, $E_7$, and $E_8$ groups.  For example, the nodes $\mu =0\dots
r=$rank$(G)$ of the extended Dynkin diagram of the $ADE$ group $G$
correspond to the irreducible representations $R_\mu$ of $\Gamma _G$,
with $|R_\mu|=n_\mu$, the Dynkin indices; therefore $|\Gamma _G|=n_\mu
n_\mu$ (throughout greek indices run from $0\dots r$ and repeated
indices are summed).  Also, $R_Q\times R_\mu = a_{\mu \nu }R_\nu$
where $R_Q$ is the fundamental two dimensional representation given by
$\Gamma _G\subset SU(2)$ and $a_{\mu
\nu }$ is one if nodes $\mu$ and $\nu$ are linked in the extended
Dynkin diagram and zero otherwise.  The character of this relation
is
\eqn\charp{\chi _Q (g)\chi _\mu (g) =a_{\mu \nu}\chi _\nu
(g),} for all $g\in \Gamma _G$.  In particular, for $g =1$, this gives
$\widetilde C_{\mu \nu}n_\nu =0$, where $\widetilde C_{\mu
\nu}=2\delta _{\mu \nu}-a_{\mu \nu}$ is the extended Cartan matrix;
the Dynkin indices $n_\mu$ give the only null vector of $\widetilde
C_{\mu \nu}$.

The ALE space $X\cong \IC ^2/\Gamma _G$ has $r$ non-trivial two cycles
$\Sigma _i$, $i=1\dots r$, generating $H_2$.  $X$ is hyper-Kahler with
a triple of Kahler forms $\vec \omega$ and
\eqn\blowup{\int _{\Sigma _i}\vec \omega =\vec \zeta _i,}
with $\vec \zeta _i$, $i=1\dots r$, the $3r$ blowing up parameters.

The ALE spaces have a hyper-Kahler quotient construction \kron;
i.e. there are supersymmetric gauge theories (with 8 super-charges)
which have the ALE spaces as their Higgs branches.  The theories have
gauge group $\prod _{\mu =0}^rU(n_\mu )/U(1)$, where the $U(1)$ in the
quotient is the overall $U(1)$.  The matter content of the gauge
theory is given by hypermultiplets in the representations $\half
\oplus_{\mu \nu=0}^r a_{\mu \nu} (\fund _\mu , \overline {\fund
}_\nu)$, where the index labels the gauge group and the $\half$ is to
avoid a double counting.  Throughout, $n_\mu$ and $a_{\mu \nu}$ are
defined as above.  The $3r$ blowing parameters $\vec \zeta _i$ enter as
the $r$ \FI\ parameters which can be included for the $r$ $U(1)$
factors in the gauge group \kron.

The above gauge theories arise physically, for example in $D0$ branes
(of IIA) or $D1$ branes (of IIB), which probe the geometry of the ALE
spaces \refs{\dm, \cjrm \mdegs}.

\subsec{Instantons on ALE spaces}

On flat $\IR ^4$, instantons in an arbitrary gauge group $\G$ (not to
be confused with the ADE group $G$ in $\IC ^2 /\Gamma _G$) are
topologically specified by the instanton number $K$, which is a
non-negative integer associated with $\pi _3(\G )\cong \IZ$.  The
moduli space $\mI$ of instantons with a given $K$ is a hyper-Kahler
space with dimension in hypermultiplet (i.e quaternionic) units (one
quarter of the real dimension) given, using the index theorem of \AS,
by $\dim (\mI) =C_2(\G )K$, where $\C _2(\G)$ is the dual Coxeter
number (index of the adjoint; e.g. $C_2 =N$ for $SU(N)$ or $N-2$ for
$SO(N)$).

When the usual $\IR ^4$ is replaced with an ALE space $X\cong \IC ^2
/\Gamma _G$, instantons are specified by certain topological data in
addition to the integer $K$.  An instanton gauge connection is
asymptotically flat and thus usually trivial since the asymptotic
space $X_\infty$ surrounding the instanton usually has trivial $\pi
_1$.  However, on the ALE space, $\pi _1(X_{\infty})=\Gamma _G$ and
thus there can be non-trivial Wilson lines at infinity, leading to
non-trivial group elements $\rho _\infty \in \G$, representing $\Gamma
_G$ in the gauge group.  Thus, in addition to $K$, the instanton is
topologically characterized by integers $w_\mu$ in $\rho _\infty
=\oplus _\mu w_\mu R_\mu$, giving the representation of $\Gamma _G$ in
$\G$ in terms of the irreps $R_\mu$.  When the gauge group $\G$ has an
Abelian factor, additional topological data is needed to specify the
instanton: because $X$ has non-trivial two cycles $\Sigma _i$,
$i=1\dots r=$rank$(G)$, there can be non-trivial first Chern classes
$u_i=\int _{\Sigma _i}(\Tr F/2\pi)$.  For simplicity, and because we
are chiefly in the case of $\G =SO(N)$ where $\Tr F=0$, we will set
the $u_i=0$ in this section.

The instanton number (second Chern class) with physical data $K$ and
$\rho _\infty$ is
\eqn\iwsu{I=K+{1\over \kappa |\Gamma _G|} G_{\mu \nu}n_\mu w_\nu,}
where $\kappa$ is the index of the embedding of $SU(2)$ in the gauge
group $\G$ (e.g. $\kappa =1$ for $SU(N)$ and $\kappa =2$ for $SO(N)$)
and $G_{\mu \nu}$ is defined by
\eqn\gdeff{G_{0\nu}\equiv G_{\mu 0}\equiv 0, \qquad 
G_{ij}\equiv C^{-1}_{ij}\quad i,j \neq 0,} with $C^{-1}$ the inverse
Cartan matrix of the $ADE$ gauge group $G$.  Explicit expressions for
the $G_{\mu \nu}$ for all gauge groups can be found, for example, in
table 7 of \slansky.  $G_{\mu \nu}$ inverts the extended Cartan matrix
$\widetilde C_{\mu \nu}$ up to its null vector $n_\mu$:
\eqn\CABG{\widetilde C_{\mu \nu}A_\nu=B_\mu\qquad  \rightarrow \qquad 
n\cdot B=0\quad \hbox{and}\quad A_\mu =\alpha n_\mu +G_{\mu
\nu}B_\nu,} where $\alpha$ is arbitrary.

The moduli space of instantons on the ALE space is again hyper-Kahler,
with dimension which can be determined by the index theorem of
\APS\ for manifolds with boundary (in this case at infinity).  In
hyper-multiplet units, the dimension is given by 
\eqn\dimIb{\dim (\mI)=C_2(\G )I+{1\over 2|\Gamma _G|}\sum _{g\neq 1} 
{\chi _{Ad}(\rho _\infty)(g)-|\G |\over 2-\chi _Q(g)},}
where the last term is the eta invariant, with the sum over $g\in
\Gamma _G$, omitting the identity $g=1$.  In computing the eta
invariant, it will be useful to define 
\eqn\Xis{X_{\mu \nu}\equiv {1\over 2|\Gamma _G |}\sum _{g\neq
1}{\chi _\mu (g) \overline{\chi _\nu(g )}-n_\mu n_\nu
\over 2-\chi _Q(g)}=\half G_{\mu \nu}-{1\over 2|\Gamma
_G|}(G_{\mu \sigma }n_\sigma n_\nu+G_{\nu \sigma}n_\sigma n_\mu).}
The identity in \Xis\ follows from \CABG\ and
\eqn\cxeqn{\widetilde C_{\mu \sigma}X_{\sigma \nu}={1\over 2|\Gamma
_G|}(|\Gamma _G|\delta _{\mu \nu}-n_\mu n_\nu),} which follows easily
{}from \charp\ and character orthogonality.

\newsec{Type II: $U(N)$ instantons on ALE spaces.}

\subsec{General aspects}

$N$ coincident type II $p+4$ branes have a supersymmetric $U(N)$ gauge
theory living in their world-volume \wbound.  $K$ coincident type II
$p$ branes living inside of this system have a supersymmetric (with
eight super-charges) $U(K)$ gauge theory living in their world volume
with $N$ matter hypermultiplets in the ${\bf K}$ and one in the
adjoint ${\bf K^2}$.  This configuration of branes has the
interpretation as $K$ $U(N)$ instantons and, indeed, this $U(K)$
world-volume theory has a Higgs branch which gives the hyper-Kahler
quotient construction of the moduli space of $K$ $U(N)$ instantons
\mrdbranes. 

Consider now the situation where the $p$ branes are on an ALE space
$\IC ^2/\Gamma _G$.  The resulting world-volume theory should have a
Higgs branch which is isomorphic to the moduli space of $U(N)$
instantons on the ALE space.  This moduli space does indeed have a
hyper-Kahler quotient construction \kn, i.e. a realization as the
Higgs branch of a supersymmetric gauge theory with 8 supercharges.  It
is natural to expect that these theories arise in the world-volume of
the $p$ branes on the ALE space; this was shown for the $\IC ^2/\IZ
_{r+1}$ orbifolds in \dm.   

As discussed in the previous section, on an ALE space an instanton can
have a non-trivial $\rho _\infty \in U(N)$ representing $\Gamma _G$:
$\rho _\infty = \oplus _\mu w_\mu R_\mu$, where the integers $w_\mu$
must satisfy $w\cdot n\equiv w_\mu n_\mu =N$ in order to have $\rho
_\infty \in U(N)$.  The group element $\rho _\infty$ breaks
$U(N)\rightarrow \prod _{\mu =0}^r U(w_\mu )$.  In addition, there can
be non-trivial first Chern classes $u_i=\int _{\Sigma _i}(\Tr
F/2\pi)$, $i=1\dots r$.  As discussed in \jbki, the $u_i$ are related
to the integral of the NSNS $B$ field by
\eqn\uBrel{u_i=N\int _{\Sigma _i}B.}

The theory of \kn\ has gauge group $\prod _{\mu =0}^r U(v_\mu )$, with
matter content consisting of $w_\mu$ hypermultiplets in the $\fund
_\mu$ and hypermultiplets in the $\half \oplus _{\mu\nu} a_{\mu
\nu}(\fund _\mu ,
\overline{\fund _\nu})$, where the subscripts label the gauge group
(and the $\half$ is to avoid double counting).  The $v_\mu$ are
determined in terms of the physical data by
\eqn\vwusu{\widetilde C_{\mu \nu}v_\nu =w_\mu -u_\mu;}
$u_{\mu >0}$ are the first Chern classes and $u_0$ is determined by
$n\cdot u=n\cdot w$, which follows from \vwusu.  Of course it is
physically necessary that the small instanton gauge group is
generically completely broken on the Higgs branch; this is indeed the
case given the relation \vwusu\ and that all $u_\mu \geq 0$.  The
dimension of the Higgs branch $\mH$ in hypermultiplet (i.e
quaternionic) units (one quarter of the real dimension) is given by
\eqn\dimHsu{\dim (\mH)=v_\mu w_\mu -\half \widetilde C_{\mu \nu}v_\mu
v_\nu= \half v_\mu (w_\mu + u_\mu ).}

Note that there are $r+1$ $U(1)$ factors in the gauge group.  The
diagonally embedded $U(1)$ factor has no charged matter and thus
decouples.  Including
\FI\ terms $\vec \zeta _i$, $i=1\dots r$, 
for the remaining $r$ $U(1)$ factors gives the dependence of the
moduli space of instantons on the $3r$ blowing up modes of the ALE
space.  We emphasize again that the construction of
\kn\ applies {\it only} for gauge group $U(N)$.

We now introduce some formulae which we will find analogs for $SO(N)$
instantons; for this purpose it suffices to restrict our attention to
$u_{\mu >0}=0$.  The instanton number (second Chern class) with
physical data $K$ and $\rho _\infty$ is given by \iwsu\ with $\kappa
=1$; thus
\eqn\vtsu{I={v_\mu n_\mu \over |\Gamma _G|},}
which, as expected, gives the instanton number as the number of $p$
branes on the orbifold. The dimension of the moduli space $\mI$ of
instantons is given by
\dimIb\ with $C_2(U(N))=N$ and $\chi _{Ad}(\rho _\infty)(g)=
w_\mu w_\nu \chi _\mu (g)\overline {\chi _\nu (g)}$.  Thus 
\eqn\dimIe{\dim (\mI)=NI+ w_\mu w_\nu X_{\mu \nu},} 
with $X_{\mu \nu}$ defined as in
\Xis.  Using \vtsu, \vwusu, and \cxeqn, it is easily verified that $\dim
(\mI )$, given by \dimIe, agrees with $\dim (\mH )$, given by \dimHsu,
giving a check of the equivalence $\mH\cong
\mI$ proven in \kn\ for $U(N)$ instantons.

\subsec{Orbifold constructions}

The above gauge theories can be directly obtained from an orbifold
construction of the theory on $\IC ^2/\Gamma _G$; for the case of
$\Gamma _G=\IZ _{r+1}$ this was shown in \dm.  As discussed in
\jbki, the relation \vwusu\ arises from a tadpole consistency
condition.  We take $\Gamma _G$ to act on the $p$ brane and $p+4$
brane Chan-Paton factors as
\eqn\typeiicp{\gamma _{g,p}=\oplus _\mu R_\mu (g) \otimes I_{v_\mu},
\qquad \gamma _{g,p+4}=\oplus _\mu R_\mu (g)\otimes I_{w_\mu},}
where $I_n$ is the $n\times n$ identity matrix.  Following the
discussion in \jbki, the relevant tadpole equations are those twisted
by $g\neq 1$, which are
\eqn\typeiitt{{1\over 2-\chi _Q(g)}\left( w_\mu \chi _\mu (g)-(2-\chi
_Q(g)) v_\mu \chi _\mu (g)\right) ^2=0.} These can be put in the form
\vwusu, though with a specific value for the $u_\mu$: $u_\mu =Nn_\mu
/|\Gamma _G|$.  As discussed in \jbki, this means that the orbifold
construction gives a specific value for the integral of the NSNS $B$
field over the $\Sigma _i$
\eqn\bint{\int _{\Sigma _i}B={n_i\over |\Gamma _G|}, \qquad i=1\dots
r.}  The fact that the $B$ field is restricted to a particular value
in the weakly coupled orbifold construction is standard, as in
\aspinB.  The particular value \bint\ agrees (only) for the $A_r$
cases with the result of \mdegs, where the result was obtained by
requiring $D0$ branes to have mass $1/C_2(G)g_s$, where $C_2(G)=\sum
_\mu n_\mu$.

\newsec{Type IIB five-branes near a $\IC ^2/\Gamma _G$ singularity}

$K$ type IIB five-branes at a point in $\IR ^4$ have a $6d$ ${\cal
N}=(1,1)$ supersymmetric gauge theory, with gauge group $U(K)$, living
in their world-volume.  This theory is anomaly free with arbitrary
gauge coupling $g$ and is IR free.  Now consider the situation when
the five-branes sit not at a generic point in $\IR ^4$ but at a $\IC
^2 /\Gamma _G$ orbifold singularity.  Much as in \dm\ and the previous
section, though with $p=5$ and without $p+4$ branes, the 6d
world-volume theory becomes a ${\cal N}=(1,0)$ supersymmetric gauge
theory with gauge group
\eqn\iibgg{\prod _{\mu =0}^r U(Kn_\mu )} 
and matter multiplets in representations $\half \oplus _{\mu
\nu}a_{\mu\nu}(\fund _\mu , \overline {\fund _\nu})$.  In addition,
there are $r$ ${\cal N}=(1,0)$ hypermultiplets and $r$ ${\cal
N}=(1,0)$ tensor multiplets (which combine into $r$ ${\cal N}=(2,0)$
matter multiplets), coming from reducing the 10d two-form and
four-form potentials down to 6d on the $r$ cycles which generate $H_2$
of $\IC ^2/\Gamma _G$.

The above world-volume theory must be free of 6d gauge anomalies.
Writing the gauge group more generally as $\prod _{\mu =0}^r
U(v_\mu)$, the 6d gauge anomaly is 
\eqn\Aiibis{\cA = \widetilde C_{\mu \nu}v_\nu \tr F_\mu ^4
+3\widetilde C_{\mu \nu}\tr F_\mu ^2 \tr F_\nu ^2.}  For $v_\mu
=Kn_\mu$ the deadly $\tr F_\mu ^4$ irreducible anomaly terms vanish
since $\widetilde C_{\mu \nu}n_\nu =0$.  The remaining reducible
anomaly in \Aiibis\ can be cancelled with $P$ tensor multiplets
provided it can be written as $\cA = \sum _{i=1}^P(A_{i\mu} \tr F_\mu
^2 )^2$ with $A_{i\mu}$ real.  The cancellation occurs by coupling the
scalar fields $\Phi _i$ of the tensor multiplets to the gauge fields
via the interactions
\eqn\tfint{\sum _{i=1}^PA_{i\mu}\Phi _i\tr F_\mu ^2 \qquad \rightarrow
\qquad  g_{\mu , eff} ^{-2}(\Phi
)=g^{-2}_{\mu ,cl}+\sum _iA_{i\mu}\Phi _i.}  

Thus, to cancel the remaining anomaly in \Aiibis, we want to write
$\cA =3\widetilde C_{\mu \nu}x_\mu x_\nu$, where we define $x_\mu
\equiv
\tr F_\mu ^2$, as a sum of squares.  This is done for $A_r$ as
\eqn\Aas{\f{1}{3}\cA=\widetilde C_{\mu \nu}x_\mu x_\nu =
\sum _{\mu =0}^{r}(x_\mu -x_{\mu +1})^2, \qquad x_{r+1}\equiv
x_0.}
For $D_r$ this is done using 
\eqn\Das{\eqalign{\f{1}{6}\cA =\half \widetilde C_{\mu \nu}x_\mu x_\nu
&=(x_0-\half x_2)^2+ (x_1-\half x_2)^2+\half \sum
_{i=2}^{r-3}(x_i-x_{i+1})^2\cr &+(\half x_{r-2}-x_{r-1})^2+(\half
x_{r-2}-x_r)^2.}} For $E_6$, this is done using 
\eqn\Evias{\eqalign{\f{1}{6}\cA =\half \widetilde C_{\mu \nu}x_\mu
x_\nu &=(x_1-\half
x_2)^2+(x_0-\half x_6)^2+(x_5-\half x_4)^2+\f{3}{4} (x_2-{2\over
3}x_3)^2\cr &+\f{3}{4}(x_6-{2\over 3}x_3)^2+{3\over 4}(x_4-{2\over
3}x_3)^2.}} For $E_7$, we have
\eqn\Eviias{\eqalign{\f{1}{6}\cA =\half \widetilde C_{\mu \nu} x_\mu
x_\nu 
&=(x_0-\half x_1)^2 +\f{3}{4} (x_1-{2\over 3}x_2)^2+{2\over
3}(x_2-{3\over 4}x_3)^2+\cr
&+(x_6-\half x_5)^2+\f{3}{4} (x_5-{2\over 3}x_4)^2+{2\over
3}(x_4-{3\over 4}x_3)^2 +(x_7-\half x_3)^2.}}
For $E_8$, we have
\eqn\Eviiias{\eqalign{\f{1}{6} \cA =\half \widetilde C_{\mu \nu} x_\mu
x_\nu &=(x_1-\half x_2)^2 +\f{3}{4} (x_2-{2\over 3}x_3)^2+(x_8-\half
x_3)^2 +(x_0-\half x_7)^2\cr &+ \f{3}{5}(x_4-{5\over 6}x_3)^2+ {5\over
8}(x_5-{4\over 5}x_4)^2+{2\over 3}(x_6-{3\over 4}x_5)^2+{3\over
4}(x_7-{2\over 3}x_6)^2.}}  Throughout we label\foot{With this
labeling, the Dynkin indices are: For $A_r$, all $n_\mu =1$. For
$D_r$, $n_\mu =1$ for $\mu =0,1, r-1,r$ and $n_\mu =2$ for $2\leq \mu
\leq r-2$. For $E_6$, $n_\mu =1$ for $\mu =0,1,5$; $n_\mu =2$ for $\mu
=2,4,6$ and $n_\mu =3$ for $\mu =3$. For $E_7$, $n_\mu =1$ for $\mu
=0,6$; $n_\mu =2$ for $\mu =1,5,7$; $n_\mu =3$ for $\mu =2,4$ and
$n_\mu =4$ for $\mu =3$. For $E_8$, $n_\mu =1$ for $\mu =0$; $n_\mu
=2$ for $\mu =1,7$; $n_\mu =3$ for $\mu =6,8$; $n_\mu =4$ for $\mu
=2,5$; $n_\mu =5$ for $\mu =4$ and $n_\mu =6$ for $\mu =3$.}  the
nodes of the extended Dynkin diagrams as in table 16 of \slansky.  In
each of the above cases, $\cA$ is a sum of $r$ squares.  Thus the
reducible anomaly can be cancelled in every case with $r$ tensor
multiplets.  This is perfect because, as discussed above, the theory
indeed has $r$ tensor multiplets.  These $r$ tensors must couple to
the gauge fields in \iibgg\ according to
\tfint\ with \Aas - \Eviiias.

The diagonal $U(1)$ factor in $\prod _{\mu =0}^rU(Kn_\mu)$ has no
charged matter and decouples.  The other $r$ $U(1)$ factors do have
charged matter and are thus anomalous.  As in the discussion in
\refs{\dm, \Betal}, $r$ hypermultiplets are needed to cancel these
anomalies.  This again is perfect because, as discussed above, we have
$r$ hypermultiplets, corresponding to the blowing up modes of the $\IC
^2 /\Gamma _G$ singularity.  They pair up with the
$U(1)$ gauge fields to give them a mass and their expectation values
effectively become \FI\ parameters in the gauge group.   As
mentioned in sect. 2.1, the blowing up modes are also realized as \FI\
parameters in the analogous $\prod U(n_i)/U(1)$ hyper-Kahler quotient
construction of \kron\ of the ALE geometry.  This direct
correspondence makes it obvious that our gauge theory \iibgg\ will be
properly behave under smoothing the ALE space: the \FI\ terms will
Higgs \iibgg\ to exactly the right analog for the smoothed ALE space.
The cancellation of the $U(1)$ anomalies using the $r$
hyper-multiplets and the cancellation of the reducible $\tr F_\mu
^2\tr F_\nu ^2$ anomalies using the $r$ tensor multiplets can be
regarded as being related by a remnant of the larger supersymmetry of
the full type IIB theory.

The theories \iibgg\ have a Higgs branch associated with giving
expectation values to the matter fields in the $\half \oplus _{\mu
\nu}a_{\mu \nu}(\fund _\mu ,
\overline {\fund _\nu})$.  On this branch,
the gauge group is broken according to $K\rightarrow K-R$, for any
$R=1\dots K$, along with an extra, decoupled, unbroken $U(R)_D$
subgroup.  The $U(R)_D$ theory has a hypermultiplet in the adjoint
representation and is diagonally embedded in all of the groups in
\iibgg.  The index of the embedding of $U(R)_D$ in each $U(Kn_\mu)$
gauge group is $n_\mu$ and thus $U(R)_D$ has a gauge coupling
\eqn\iibdgc{g_D^{-2}=\sum _\mu n_\mu g_\mu ^{-2}.}  
The gauge couplings $g_\mu ^{-2}$ depend on the tensor multiplet
expectation values according to \tfint\ with \Aas\ - \Eviiias.  In
each case, though, it follows from $\widetilde C_{\mu \nu} n_\nu =0$
that the linear combination \iibdgc\ is independent of the $\ev{\Phi
_i}$, $i=1\dots r$, tensor multiplet expectation values.  

The interpretation of the above part of the Higgs branch is that it
corresponds to moving $R$ of the IIB five-branes away from the $\IC ^2
/\Gamma _G$ singularity.  The $U(R)_D$ theory with a hypermultiplet in
the adjoint and constant gauge coupling is exactly the expected ${\cal
N}=(1,1)$ theory for the five-branes away from the ALE singularity.
Also, the $D$ terms for the single hypermultiplet mode labeling motion
along this flat direction can always be mapped to exactly those of the
theory of
\kron, described in sect. 2.1, which gives the $\IC ^2 /\Gamma _G$ ALE
space.  In other words, the $R$ five-branes, which should be able to
wander around any point on $\IC ^2 /\Gamma _G$, indeed can.

The theory \iibgg\ for type IIB branes at the $\IC ^2/\Gamma _G$
singularity can be given an explicit orbifold construction, following
\dm.  The $r$ ${\cal N}=(1,0)$ hyper and tensor multiplets come from
the $r$ twisted conjugacy classes of $\Gamma _G$. Following the
discussion in \jbki, the twisted tadpoles reproduce the spacetime
anomaly \Aiibis, which led to $v_\mu =Kn_\mu$.

As the simplest example, consider $K$ IIB five-branes at a $\IC ^2
/\IZ _2$ $A_1$ ALE singularity.  The world-volume theory is
$U(K)\times U(K)$ with hypermultiplets in the $(\fund ,
\overline{\fund})\oplus (\overline{\fund},
\fund )$.  These theories were introduced in \Sanom\ as a class of
theories which satisfy the anomaly cancellation condition when coupled
to gravity.  Here we see that the gauge anomaly can be cancelled even
with gravity decoupled by including an extra tensor multiplet.

When the $\IC ^2 /\IZ _2$ singularity is not blown up and thus the
$U(K)\times U(K)$ theory has no \FI\ term, there is a direction in the
Higgs branch where the matter fields $Q_1$ in the $(\fund ,
\overline{\fund})$, and $Q_2$ in the $(\overline{\fund}, \fund)$ have
expectation values $\ev{Q_f}=diag(
\ev{E_f}\times 1_{R}, 0\times 1_{K-R})$ where $I_{R}$ is the $R\times
R$ identity matrix, for any $1\leq R\leq K$, and $E_f$, $f=1,2$, are
hypermultiplets.  This is a $D$-flat direction provided the $\ev{E_f}$
satisfy the $D$ term equations for $U(1)$ with two electrons, giving
precisely the construction of \kron, reviewed in sect. 2.1, for the
singular $\IC ^2 /\IZ _2$ ALE space.  The Higgs branch associated with
the above $\ev{Q_f}$ is thus the mode for moving $R$ five-branes away
{}from the singularity.  When the $\IC ^2/\IZ _2$ singularity is
resolved and thus the $U(K)\times U(K)$ theory has \FI\ term, the
above Higgs branch only occurs for $R=K$, with the $\ev{E_f}$
satisfying the $D$ term equations for $U(1)$ with two electrons and
non-zero \FI\ term. Since $\ev{E_f}\neq 0$, we find that $U(K)\times
U(K)$ is necessarily Higgsed to $U(K)_D$ at a scale set by the blowing
up parameter.  Thus, as should be expected, we necessarily obtain the
the standard ${\cal N}=(1,1)$ supersymmetric $U(K)$ on the five-branes
when the singularity is resolved.  There is a similar analysis for the
theories \iibgg\ for general $\IC ^2/\Gamma _G$ singularities.

Following \refs{\sdfp, \obrane}, there can be a non-trivial 6d RG
fixed point at the origin of the Coulomb branch provided all $g_{\mu,
eff}^{-2} (\Phi )$ are non-negative along some entire ``Coulomb
wedge'' of allowed $\Phi _i$, $i=1\dots r$.  This should be true even
in the limit when all $g^{-2}_{\mu , cl}\rightarrow 0$ in order to
obtain a RG fixed point theory at the origin.  However since \iibdgc\
is a constant independent of the $\ev{\Phi _i}$ on the Coulomb branch,
at least one of the $g_{\mu, eff}^{-2}$ must become negative for large
$\ev{\Phi _i}$.  This corresponds to the fact that the $U(R)_D$
subgroup is always IR free.  As in \obrane, we can always take the
$U(K)$ corresponding to the extended Dynkin node $\mu =0$ to be the IR
free theory, which means that this gauge group is un-gauged in the IR
limit.  It is then possible to choose a Coulomb wedge so that the
remaining gauge groups in \iibgg\ all have $g_{\mu, eff} ^{-2}(\Phi
)\geq 0$ along the entire wedge even in the $g_{\mu
,cl}^{-2}\rightarrow 0$ limit.

Thus for any $\IC ^2/\Gamma _G$, and for every $K$, these theories
give 6d non-trivial RG fixed points with $r$ tensor multiplets and
gauge group $\prod_{\mu =1}^rSU(Kn_\mu)$, where the $U(1)$s in \iibgg\
have been eliminated, as discussed above, by the anomaly cancellation
mechanism, and the $\mu =0$ node gives a global rather than local
$U(K)$ symmetry.

The simplest example, based on $K$ IIB five-branes at a $\IC ^2/\IZ
_2$ singularity, thus gives a 6d ${\cal N}=(1,0)$ supersymmetric
theory with gauge group $SU(K)$, $2K$ matter fields in the $\fund$,
and a tensor multiplet.  There is a non-trivial RG fixed point, for
all $K$, at the point where the tensor multiplet has zero expectation
value.  Another string theory realization of these theories was given
in \bervaf.

\newsec{$SO(N)$ instantons on ALE spaces}

As in sect. 2.2, we can have Wilson line group elements $\rho _\infty
\in SO(N)$ representing $\Gamma _G$.  Following the discussion in
\Betal, the type I or heterotic gauge group is really $Spin(32)/\IZ
_2$, where the $\IZ _2$ is generated by the element $w$ in the center
of $Spin(32)$ which acts as $-1$ on the vector, $-1$ on the spinor of
negative chirality, and $+1$ on the spinor of positive
chirality. Because only representations with $w=1$ are in the
$Spin(32)/\IZ _2$ string theory, the identity element $e\in \Gamma _G$
can be mapped to either the element $1$ or $w$ in $Spin(32)$.  Here we
will only consider the case $e\rightarrow 1$, which is the case with
possible vector structure.

The basic difference from the $U(N)$ case is that $SO(N)$ is a real
group.  Starting from $U(N)$, $SO(N)$ can be regarded as being
obtained by modding out by the complex conjugation operation $*$; for
example the ${\bf N}$ and ${\bf \overline N}$ representations of
$U(N)$ are both mapped to the ${\bf N}$ representation of $SO(N)$.
Properly understanding this operation of $*$, it will be fairly
straightforward to modify the results reviewed in sect. 3 to the
$SO(N)$ case.  In the orientifold discussion of sect. 6, modding out
by the operation $*$ is, of course, the orientifold process.

It will be useful in what follows to group the set of all $\{\mu
=0\dots r\}$ into the subsets as $\{\mu \}=\R\oplus \P\oplus
\C\oplus
\overline\C$; $\Gamma _G$ representations $R_\mu$ with $\mu \in \R$
are real; representations $R_\mu$ with $\mu\in \P$ are pseudo-real;
representations $R_\mu$ with $\mu \in \C$ are complex, with the
complex conjugate representation $\overline{R_\mu}=R_{\overline \mu}$,
with $\overline \mu \in \overline C$.  In terms of the $SU(2)$ which
contains $\Gamma _G$, the real representations have integer spin and
the central element $C$ of $SU(2)$ is represented by $C=1$, while the
pseudo-real representations have half-integer spin and the central
element is represented as $C=-1$.  Clearly, one dimensional
representations are either real or complex, not pseudo-real.  Using
the connection between $\Gamma _G$ and the $ADE$ extended Dynkin
diagrams, since $\chi _Q\times R _\mu =a_{\mu \nu}R_\nu$ and $\chi _Q$
has spin $j=\half$, links of the diagram connect nodes with opposite
values of $C$.  The decompositions of the representations are as
follows (with sets not listed empty).  For $A_r$ with $r$ odd,
$\R=\{0,~\half (r+1)\}$, $\C=\{1, \dots , \half (r-1)\}$, $\overline
C=\{\half (r+3), \dots , r\}$.  For $A_r$ with $r$ even, $\R=\{0\}$,
$\C=\{1, \dots, \half r\}$, $\overline C=\{\half r+1, \dots, r\}$.
For $D_r$, with $r$ odd, $\R=\{0,~ 1,~ 2s+1|~s=1, \dots , \half
(r-3)\}$, $\P=\{2s| ~s=1, \dots , \half (r-3)\}$, $\C=\{r-1\}$,
$\overline C=\{r\}$.  For $D_r$ with $r$ even, $\R=\{0,~1,~r-1,~r,~
2s+1|~s=1\dots \half (r-4)\}$, $\P=\{2s|~s=1, 
\dots ,  \half (r-2)\}$.  For $E_6$, $\R=\{0,~3\}$, $\P=\{6\}$,
$\C=\{1,~2\}$, $\overline C=\{4,~5\}$.  For $E_7$,
$\R=\{0,~2,~4,~6,~7\}$, $\P=\{1,~3,~5\}$.  For $E_8$,
$\R=\{0,~2,~4,~6,~8\}$, $\P=\{1,~3,~5,~7\}$.  

A useful quantity here is the ``Frobenius-Schur indicator,'' which is
defined (see, e.g. \collins ) to be $+1$ for real representations;
$-1$ for representations which are not real but whose characters are
real, i.e. pseudo-real representations; and $0$ for representations
with complex character.  We thus define $S_\mu \equiv +1$ for $\mu \in
\R$, $S_\mu \equiv -1$ for $\mu \in \P$, and $S_\mu \equiv 0$ for $\mu
\in \C$ or $\overline \C$.  By a theorem discussed e.g. in \collins,
\eqn\Seqn{S_\mu ={1\over |\Gamma _G|}\sum _{g\in \Gamma _G}\chi _\mu
(g^2).}

The representation $\rho _\infty$ of $\Gamma _G$ in $SO(N)$ can be
written as $\rho _\infty =\oplus _{\mu}w_\mu R_\mu$, where $n_\mu
w_\mu =N$.  In order for $\rho _\infty$ to be real, for every copy of a
complex representation $R_\mu$ with $\mu \in \C$, there has to be a
copy of the complex conjugate representation
$\overline{R_\mu}=R_{\overline \mu}$, with $\overline \mu \in
\overline \C$: i.e $w_\mu =w_{\overline \mu}$.  The pseudo-real
representations also must satisfy a condition to fit in $SO(N)$:
because they are representations of $SU(2)$ rather than $SO(3)$, they
need to be embedded in an $SO(4)\cong SU(2)\times SU(2)$ subgroup of
$SO(N)$.  This means that $w_\mu$ must be even for $\mu \in \P$.
Finally, the real representations are already representations of
$SO(3)$ and thus fit into $SO(N)$ without the need for any such
doubling.  The non-trivial $\rho _\infty$ breaks $SO(N)$ to a subgroup
as
\eqn\sobreak{SO(N)\rightarrow \prod _{\mu \in \R}SO(w_\mu )\times 
\prod_{\mu \in \P}Sp(\half w_\mu)\times \prod _{\mu \in \C}U(w_\mu).}

The relevant ``small instanton'' theory is a supersymmetric gauge
theory (with eight super-charges) with gauge group
\eqn\sigg{\prod _{\mu \in \R}Sp(v_\mu )\times \prod _{\mu \in
\P}SO(v_\mu )\times \prod _{\mu \in \C}U(v_\mu ).}
The group of \sobreak\ arises as a global symmetry of the gauge theory
\sigg; thus the $Sp(v_\mu)$ has $w_\mu$ half-hypermultiplets in
the $\fund _\mu$, the $SO(v_\mu)$ has $\half w_\mu$ hypermultiplets in
the $\fund _\mu$, and the $U(v_\mu )$ has $w_\mu$ hypermultiplets in
the $\fund _\mu$.  In addition, there are hypermultiplets
corresponding to the links of the extended Dynkin diagram\foot{The
case of $\Gamma _G=\IZ _{2P+1}$ discussed in \refs{\dm, \obrane} is a
special case in that the group $U(v _P)$ also has a hypermultiplet in
the $\asym _P$.  In terms of the discussion here, the unique aspect of
$\Gamma _G=\IZ _{2P+1}$ is that the extended $SU(2P+1)$ Dynkin diagram
is the only case which does not admit a two coloring of the nodes,
with linked nodes of opposite colors, corresponding to whether the
corresponding representation of $SU(2)$ has integer or half-integer
spin. The results below thus apply for every case except for $\Gamma
_G=\IZ _{2P+1}$, which is covered by the analysis of \obrane.}: there
are hypermultiplets in the $\f{1}{4}
\oplus _{\mu \nu} a_{\mu \nu}(\fund _\mu , \fund _\nu )$, where the
sum runs over all $\mu =0\dots r$, with $\fund _{\overline \mu }\equiv
\fund _{\mu }$ for $\overline \mu
\in \overline \C$. In the $\f{1}{4}\oplus _{\mu \nu}$, 
one factor of $\half$ is for a double counting in $\oplus _{\mu \nu}$
and the other is because either: $\mu
\in \R$ and $\nu \in \P$ (or vice-versa) and thus the gauge group is
$Sp(v_\mu )\times SO(v_\nu)$ and the matter is in a
half-hypermultiplet of $(\fund _\mu ,\fund _\nu )$; or $\mu $ (or
$\nu$) $\in \C$, in which case the extra factor of $\half$ is for a
second double counting associated with $\overline \mu$  (or $\overline
\nu $) $\in \overline C$.

The $v_\mu$ data in \sigg\ will be related to the physical data
$w_\mu$ which specify $\rho _\infty$ by a relation of the form
\eqn\cvwso{\widetilde C_{\mu \nu} V_\nu =w_\mu -D_\mu,}
for some $D_\mu$ with $n\cdot D=n\cdot w=N$, and $V_\mu \equiv \dim
(\fund _\mu )$ (i.e. $V_\mu=2v_\mu$ for $Sp(v_\mu)$ and $V_\mu =v_\mu$
for $SO(v_\mu)$ and $U(v_\mu )$).  Repeated indices run over all $\mu
=0\dots r$, with $w_{\overline \mu}\equiv w_\mu$ and $V_{\overline
\mu} \equiv V_{\mu}$ for $\overline \mu \in \overline \C$ and $\mu \in
\C$.  Using \CABG, the solution of \cvwso\ with $v_0=K$ is 
\eqn\vsois{V_\mu =2Kn_\mu +G_{\mu \nu}(w_\nu -D_\mu).}
In analogy with results for $U(N)$ quiver gauge groups, we expect that
the gauge theory \sigg\ has a Higgs branch on which the gauge group
can be completely broken by the Higgs mechanism provided $N\geq 3$ and
all $D_\mu \geq 0$.  In this case, the dimension of the Higgs branch
is given by
\eqn\dimmhso{\dim (\mH)=\f{1}{4}V_\mu (w_\mu +D_\mu )-\half V_\mu S_\mu,}
where $S_\mu$ is the Frobenius-Schur indicator discussed above the
theorem \Seqn.

First, we argue that the Higgs branch $\mH$ of the above gauge
theory is isomorphic to the moduli space $\mI$ of $SO(N)$ instantons
on the ALE space, giving a hyper-Kahler quotient construction of this
moduli space, when \cvwso\ is satisfied and 
\eqn\Dhk{D_\mu =N\delta _{\mu 0} \qquad \hbox{(hyper-Kahler
quotient)}.}  A caveat is that $r-|\C |$ of the blowing up modes of
the ALE space can not be turned on, they are frozen at zero.  Next we
argue that the small instanton theory \sigg\ with the data
\Dhk\ can not arise physically in the world-volume of five-branes at
the ALE singularity as it would have a deadly anomaly.  There is a
simple modification of \Dhk\ for which the anomalies vanish:
\eqn\Dphy{D_\mu =16S_\mu \qquad \hbox{(physical)},}
where we define $S_\mu$ as above.  Note that the condition \Dphy\ in
\cvwso\ properly gives $N=w_\mu n_\mu =D_\mu n_\mu =16S_\mu n_\mu
=32$, where the identity $S_\mu n_\mu =2$ follows from \Seqn\ and
character orthogonality, using $n_\mu =\chi _\mu (h=1)$ and $\sum
_g\delta _{g^2,1}=2$, since there are always precisely two elements,
$g=1$ and the central element $C$ of $SU(2)$, which square to unity.
As expected, consistency of the 5-brane theory requires 32 9-branes.

The instanton number (second Chern class) with a given $\rho_\infty$
is given by \iwsu\ with $\kappa =2$, corresponding to the index of the
embedding of $SU(2)$ in $SO(N)$ (in an $SO(4)\cong SU(2)\times
SU(2)$).  Using \vsois, 
\eqn\Vnis{I={1\over 2|\Gamma _G|}(V\cdot n  +G_{\mu \nu}D_\mu n_\nu).} 
On the orientifold, $V\cdot n/ 2 |\Gamma _G|$ is
the number of five-branes.  In particular, with \Dhk\ the last term in
\Vnis\ vanishes and the result \Vnis\ gives that the instanton number
is the number of five-branes in the hyper-Kahler quotient case.

The dimension of the moduli space of $SO(N)$ instantons on the ALE
space is given by
\dimIb\ with $C_2(SO(N))=N-2$ and $\chi _{Ad}(\rho _\infty)(g)=\half
w_\mu w_\nu \chi _\mu (g)\chi _\nu (g)-\half 
w_\mu \chi _\mu (g^2)$, which follows because $\chi _{Ad}(\rho _\infty )$ 
is the antisymmetric product of $\rho _\infty\times \rho _\infty$ for
$SO(N)$.  Thus
\eqn\dimIso{\dim (\mI)=(N-2)I+\half w_\mu 
w_\nu X_{\mu \nu}- w_\mu F_\mu,}
with $X_{\mu\nu}$ defined by \Xis\ (and we used $w_{\overline
\nu }=w_\nu)$ and $F_\mu$ is defined by
\eqn\Fis{F_\mu \equiv {1\over 4|\Gamma _G|}\sum _{g\neq 1}{\chi _\mu
(g^2)- n_\mu \over 2-\chi _Q(g)}.}  To evaluate $F_\mu$, note that
\eqn\ConF{\widetilde C_{\mu \nu}F_\nu= 
{1\over 4|\Gamma _G|}\sum _{g\neq
1} \left({2-\chi _Q(g^2)\over 2-\chi _Q(g)}\right)\chi _\mu (g^2)=
\half S_\mu -{n_\mu \over
|\Gamma _G|}+{1\over 4|\Gamma _G|}\sum _g \chi _Q(g)\chi _\mu (g^2),}
where we used $2-\chi _Q(g^2)=(2-\chi _Q(g))(2+\chi _Q(g))$, which can
be seen by $\chi _Q(g)\equiv 2\cos (\theta _g)$, since then
$\chi _Q(g^2)=2\cos (2\theta _g)$, and also \Seqn.  The last sum in
\ConF\ vanishes, as can be seen by noting that $\chi (g^2)=\chi
_S(g)-\chi _A(g)$, where $S$ and $A$ are the symmetric and
antisymmetric products; it follows from this that $\chi _\mu (g^2)$
has integer spin $j$ and thus has trivial projection on $\chi _Q(g)$,
which has spin $j=\half .$  Therefore, using \CABG, we find $F_\mu 
=\half G_{\mu \nu}S_\nu -(|\Gamma _G|)^{-1}G_{\mu \nu}n_\nu$.  Using
this in \dimIso, applying various other formulae appearing above, and
comparing with \dimmhso, we find
\eqn\dmimh{\dim (\mI)=\dim (\mH) -\half G_{\mu \nu}S_\mu D_\nu  
+\f{1}{4}G_{\mu \nu}D_\mu D_\nu.}
In particular, with \Dhk, $\dim (\mI) =\dim (\mH )$, giving a check of
our conjecture that in this case $\mH \cong \mI$.  

As a further check of $\mH \cong \mI$, it can be checked that the
gauge theory \sigg\ has a flat direction in the Higgs branch,
associated with giving expectation values to the matter fields in the
$\f{1}{4}\oplus a_{\mu \nu}(\fund _\mu ,
\fund _\nu )$, which breaks the gauge group according to $V_\mu
\rightarrow V_\mu -2Rn_\mu$, for any $R=1\dots K$, with an extra,
decoupled, unbroken $Sp(R)_D$ subgroup.  The $Sp(R)_D$ has the
standard matter content of \wsmall, $16\fund _D\oplus \asym _D$, and
is diagonally embedded in all of the groups in \sigg.  The index of
the embedding of $Sp(R)_D$ in each group is $n_\mu$ and thus $Sp(R)_D$
has gauge coupling
\eqn\sprgc{g_D^{-2}=\sum _\mu n_\mu g_\mu ^{-2}.}
The interpretation of this part of the Higgs branch is that $R$ small
instantons have been moved away from the $\Gamma _G$ orbifold
singularity.  Indeed, it appears that the $D$ terms for the single
hypermultiplet Higgs mode labeling motion along this flat direction
can always be mapped to exactly those of the theory of \kron,
described in sect. 2.1, which gives the $\IC ^2/\Gamma _G$ ALE space.
In other words, the location of the center of the $R$ small $SO(32)$
instantons, which should be able to wander around on the entire $\IC
^2/\Gamma _G$ ALE space, indeed can.

Because of the $|\C |$ $U(v_\mu)$ factors in \sigg, it is possible to
turn on $|\C |$ \FI\ terms, which gives the modification of the small
instanton gauge group when the corresponding blowing up mode of the
singularity is turned on.  On the other hand, for the $SO(v_\mu)$ or
$Sp(v_\mu)$ factors in \sigg, there can be no such \FI\ term and thus
the corresponding blowing-up modes are frozen.  The number of frozen
blowing up modes is $r-|\C |$.  For $D_r$ with $r$ even, $|\C |=0$ and
thus all $r$ blowing up modes are frozen.  For $D_r$ with $r$ odd,
$|\C |=1$ and thus there are $r-1$ frozen blowing up modes.  For
$E_6$, $|\C |=2$ and thus there are 4 frozen blowing up modes.  For
$E_7$ and $E_8$, $|\C |=0$ and thus there are 7 and 8, respectively,
frozen blowing up modes.  These ``missing'' blowing up modes will be
remembered shortly.

Despite the above evidence that $\mH \cong \mI$ when $D_\mu =N\delta
_{\mu 0}$ (subject to the above caveat that some blowing up modes are
frozen), these gauge theories can {\it not} arise in the world-volume
of small $SO(N)$ instanton D5 branes sitting at the ALE singularity.
Indeed, such theories have deadly $\tr F^4$ anomalies in the
world-volume of the five-branes, which renders them inconsistent.
This sickness can not be cured by adding more fields, even gravity.
However, as in the $A_r$ case \obrane, there is a simple modification
of the relation for which this deadly anomaly vanishes.  The
six-dimensional gauge anomaly for theories of the above type with
general choices of $w_\mu$ and $v_\mu$ is given by
\eqn\ganomi{\cA =\half (\widetilde C_{\mu \nu} V_\nu
-w_\mu +16S_\mu)\tr F_\mu ^4+\f{3}{2}\widetilde
C_{\mu \nu} \tr F_\mu ^2\tr F_\nu ^2,} where $S_\mu$ is defined as
above and $F_{\overline \mu}\equiv F_\mu$ along with
$V_{\overline \mu}\equiv V_\mu$ and $w_{\overline
\mu}\equiv w_\mu$ for $\overline \mu \in \overline \C$.
Thus we conjecture that the physical gauge group living in the
world-volume of the D5 branes is a gauge theory of the above type but
with the relation \cvwso\ with \Dphy, $D_\mu =16S_\mu$.

Having cancelled the deadly $\tr F^4$ anomalies, the remaining anomaly
in \ganomi\ can be cancelled by coupling the 6d gauge theory to extra
tensor multiplets, much as in the theories discussed in the previous
section.  Using \Aas\ - \Eviias, the reducible anomaly in \ganomi\ is
the sum of $r$ square terms. However, with the identification
$x_{\overline
\mu}\equiv x_\mu$ for $\overline
\mu \in \overline C$, $\cA$ is actually the sum of 
$r-|\C|$ independent square terms.  Thus the anomaly can be cancelled
with $r-|\C|$ tensor multiplets.  This leads to effective gauge
couplings which vary over the Coulomb branch as in \tfint.  As in the
theories discussed in the previous section, it follows from
$\widetilde C_{\mu \nu}n_\nu =0$ that the linear combination \sprgc\
is constant on the Coulomb branch.  This is sensible as the $Sp(R)_D$
of $R$ small instantons away from the singularity should have a
constant gauge coupling.

Finally, the $|\C |$ $U(1)$ factors are eliminated by the anomaly
cancellation mechanism using the $|\C |$ hyper-multiplet blowing up
parameters which remain massless on the Coulomb branch.

With $D_\mu =16S_\mu$, \Vnis\ gives for the number of five-branes on
the orientifold
\eqn\Vnmod{{1\over 2|\Gamma _G|}V_\mu n_\mu =I-{8\over |\Gamma _G|}G_{\mu
\nu}n_\mu S_\nu= I-(r+1),}
Where the last identity for $G_{\mu \nu}n_\mu S_\nu$ can be verified
in every case using the explicit expressions for the $G_{\mu \nu}$ in
table 7 of \slansky.  The result
\Vnmod\ has a natural interpretation: 
$r+1$ is the Euler character of the ALE space $\IC ^2/\Gamma _{G_r}$,
which is the instanton number associated with the standard embedding,
when the gauge connection is set to equal the spin connection.  Since
five-branes carry $H$ charge, \Vnmod\ is the standard relation giving
the total (magnetic) $H$ charge\foot{Including gravity, the tensors
also couple to the nine-brane gauge groups \sobreak\ according to the
standard Green-Schwarz mechanism.  When combined with the fact that
$\int _{ALE}dH
\neq 0$ with five-branes present, this means that there is an anomaly
inflow mechanism, with net current flowing onto the brane from the
ten-dimensional world.  This is the limit of the mechanism of
\ref\jbjh{J. Blum and J. Harvey, hep-th/9310035, \np{416}{1994}{119}.}
where the instantons shrink to zero size.}  as the second Chern class
$I$ minus the Euler character. In particular, for the standard
embedding there are no five-branes.

With $D_\mu =16S_\mu$, it is no longer true that $\mH\cong \mI$.
Indeed, even their dimensions differ: as seen from \dmimh, 
\eqn\dmimhn{\dim (\mH)=\dim (\mI)-56G_{\mu \nu}S_\mu S_\nu=\dim
(\mI)-28(r-|\C |),} where the last equality can be verified in every
case using the explicit forms for the $G_{\mu \nu}$.  We are thus
missing $28(r-|\C |)$ hypermultiplet moduli associated with the gauge
bundle.  Combining these with the $r-|\ C |$ missing blowing-up modes,
we see that a transition has occurred where $29(r-|\C |)$
hypermultiplets have been traded for $r-|\C |$ tensor multiplets.
This is perfect, because $r-|\C |$ tensor multiplets is precisely what
we needed to cancel the anomaly \ganomi.

To summarize, we have presented evidence that small $SO(N)$ instantons
on a $\IC ^2/\Gamma _G$ orbifold singularity have a ``Coulomb branch''
with $r-|\C |$ extra tensor multiplets and the theories \sigg\ with
\cvwso, for $D_\mu =16S_\mu$, arising in the world-volume of the D5
branes.  Note that this Coulomb branch can only exist when all
$V_\mu$, given by
\vsois\ with $D_\mu =16S_\mu$, are non-negative, giving a lower bound
on the allowed $K$.

As a simple concrete example, consider the $D_4$ case with trivial
$\rho _\infty =1$.  The 6d brane world-volume gauge group \sigg\ is
$Sp(v_0)\times Sp(v_1)\times SO(v_2)\times Sp(v_3)\times Sp(v_4)$,
with $32$ half-hypermultiplets in the $\fund _0$ and
half-hypermultiplets in the $(\fund _\mu , \fund _2)$ for $\mu
=0,1,3,4$.  Using \vsois\ and \Dphy, $v_0=K$, $v_1=K-8$, $v_2=4K-16$,
$v_3=K-8$, $v_4=K-8$.  In addition, there are the $r-|\C|=4$ extra
tensor multiplets.  Exactly this theory arose in the analysis of
\aspin, via the $F$ theory description of compactification of the
heterotic theory on a compact $K3$, of the ``collision of two hidden
obstructors.''  It was conjectured in \aspin\ that this theory should
be related to a $D_4$ singularity.  Here we see it as a special case
of theories obtained for small $SO(32)$ instantons at arbitrary $\IC
^2/\Gamma _G$ singularities.

Following \refs{\sdfp, \obrane}, there can be a non-trivial 6d RG
fixed point at the origin of the Coulomb branch provided all $g_{\mu,
eff}^{-2} (\Phi )$, given by \tfint\ using \Das\ - \Eviiias, are
non-negative along some entire ``Coulomb wedge'' of allowed $\Phi _i$,
$i=1\dots r-|\C |$.  This should be true even in the limit when all
$g^{-2}_{\mu , cl}\rightarrow 0$ in order to obtain a RG fixed point
theory at the origin.  However, since \sprgc\ is a constant
independent of the $\ev{\Phi _i}$ on the Coulomb branch, at least one
of the $g_{\mu}^{-2}$ must become negative for large $\ev{\Phi _i}$.
This corresponds to the fact that the $Sp(R)_D$ subgroup is always IR
free.  As in \obrane, we can always take the $Sp(v_0)$ corresponding
to the extended Dynkin node to be the IR free theory, which means that
this gauge group is un-gauged in the IR limit.  It is then possible to
choose a Coulomb wedge so that the remaining gauge groups in \sigg\
all have $g_{\mu, eff} ^{-2}(\Phi )\geq 0$ along the entire wedge even
in the $g_{\mu ,cl}^{-2}\rightarrow 0$ limit.  Thus each of our
theories give an infinite family of 6d non-trivial RG fixed points,
labeled by $K$, with the $Sp(v_0)$ factor in the gauge group \sigg\ a
global rather than gauge symmetry and the $U(1)$ factors eliminated by
the anomaly cancellation mechanism.

\newsec{Orientifold construction of the theories}

Type I on the $\IC ^2/\Gamma _G$ singularity can be constructed via an
orientifold of type IIB by including $\Omega$, the element that
reverses the orientation, along with $\Gamma _G$ in the orbifold
group; $\Omega ^2=1$, and $\Omega g=g\Omega$.  We must choose matrices
$\gamma _g$ and $\gamma _\Omega$ to represent the action of $\Gamma
_G$ and $\Omega$ on the Chan-Paton factors.  The $\gamma _g$ must
represent the $\Gamma _G$ group relations up to a phase.  In addition,
there are two more algebraic consistency conditions \refs{\GP , \dm}:
\eqn\conone{\gamma_{\Omega}=\pm\gamma_{\Omega}^T}
where the upper(lower) sign is for 9(5) branes, and 
\eqn\contwo{\gamma_g\gamma_{\Omega}\gamma_g^T=\gamma_{\Omega}.}
Requiring the Chan Paton factor $\gamma _C$ corresponding to the
central element $C=-1$ of $SU(2)$ to act as $\gamma _C=1$ in the
trivial representation, there is no freedom for an extra phase in the
gamma matrices and we should expect a single possibility, with 32
nine-branes.

The action of $\Gamma _G$ on the five-brane and nine-brane Chan-Paton
factors is as in \typeiicp. The action of $\Omega$ on the five-branes
is represented as
\eqn\omegfb{\gamma _{\Omega ,5}=\oplus _{\mu\in \R}(I_{n_\mu}\otimes
J_{V_\mu})\ \oplus _{\mu
\in \P}(J_{n_\mu}\otimes I_{V_\mu })\ \oplus _{\mu \in C}(J_2
\otimes I_{V_\mu}),} where 
$J_n$, with $n$ even, is the $n\times n$ symplectic matrix:
$J_n=-J_n^T$, with $J_n^2=-I_n$.  In the last term in \omegfb, $J_2$
acts to take $\mu \in \C$ to $\overline
\mu \in \overline \C$.  Similarly, $\Omega$ acts on nine-branes as 
\eqn\omegnb{\gamma _{\Omega ,9}=\oplus _{\mu \in \R}(I_{n_\mu}\otimes
I_{w_\mu})\ \oplus _{\mu \in \P}(J_{n_\mu }\otimes J_{w_\mu})\ \oplus
_{\mu \in \C}(\sigma _1\otimes I_{w_\mu}),}
where in the last term $\sigma _1$ is a symmetric Pauli matrix which 
exchanges $\mu \in \C$ with $\overline \mu \in \overline \C$.
These yield precisely the gauge groups \sigg\ for five-branes and
\sobreak\ for nine-branes.

As always, the untwisted nine-brane tadpoles can be
cancelled by having 32 nine-branes, i.e. $w\cdot n=32$.  The tadpoles
in the sector twisted by $g \neq 1$ can be written in the form:
\eqn\sotti{{1\over 2-\chi _Q(g)}\left(w_\mu \chi _\mu (g)-(2-\chi
_Q(g))v_\mu \chi _\mu (g)-16\sum _h\delta _{g,h^2}\right) ^2.}
Using the technique discussed in \jbki, we can extract the irreducible
anomaly coefficient of $\tr F_\mu ^4$:
\eqn\irano{-|\Gamma _G|(-\widetilde C_{\mu \nu}v_\nu +w_\mu -D_\mu
)-n_\mu (32-w\cdot n),}
where
\eqn\iranD{D_\mu \equiv {16\over |\Gamma _G|}\sum _g \sum _h \chi _\mu
(g)\delta _{g,h^2}=16S_\mu,} and we used \Seqn.  The $32-w\cdot n$
term in \irano\ vanishes by virtue of the untwisted nine-brane
tadpole.  We thus find perfect agreement with the irreducible
space-time anomaly \ganomi\ of the five-brane gauge theory.  The
coefficient of $\tr F_\mu ^2\tr F_\nu ^2$ in the reducible anomaly is
similarly obtained from \sotti, using the technique of \jbki, to be
proportional to $\widetilde C_{\mu
\alpha }\sum _{g\neq 1}\chi _\alpha (g)\chi _\nu (g)=|\Gamma
_G|\widetilde C_{\mu \overline{\nu}}$, which agrees with the reducible
anomaly in \ganomi\ (since $F_{\overline{\nu}} =F_\nu$).

To calculate the closed string spectrum, we note that the closed
string partition function and Klein bottle splits into terms that are
the Abelian groups representing the conjugacy classes.  For all
$\Gamma _G$ we can apply the result of \jbki\ for the case with vector
structure: we get a tensor from each twisted conjugacy class which is
self-conjugate (i.e. contains its inverse) whereas, if it is conjugate
to another twisted sector conjugacy class, we get a tensor and a
hypermultiplet from the two twisted sectors.  The number of conjugacy
classes is the same as the number of irreps, which is $r+1$ for the
$\Gamma _{G_r}$.  One of these is the identity, so there are $r$
twisted sector conjugacy classes.  Conjugacy classes containing their
inverse have real character and thus there are $r-2|\C |$
self-conjugate classes and $|\C |$ remaining conjugate pairs.  Thus we
get $r-|\C |$ tensor multiplets and $|\C |$ hypermultiplet blowing up
modes from the closed string twisted sectors.  In agreement with the
discussion in the previous section, we are missing $r-|\C |$
hypermultiplet blowing up modes (\FI\ parameters) and instead have
$r-|\C |$ tensor multiplets.

Finally, as discussed in \jbki, the $r-|\C |$ tensor multiplets will
have the correct couplings \tfint\ to the gauge fields to cancel the
reducible $\widetilde C_{\mu \nu}\tr F_\mu ^2\tr F_\nu ^2$ anomaly.
Similarly, the $|\C |$ hypermultiplets have the correct couplings to
cancel the $U(1)$ anomalies in the five-brane gauge group \sigg.

The above, perturbative, orientifold analysis applies far out along
the ``Coulomb branch,'' where the tensors have large expectation value
and the theory is weakly coupled.  As discussed in \jbki, the
orientifold construction gives an intuitive picture for how the extra
tensors of the Coulomb branch arise: starting from IIB on the ALE
space and orientifolding to obtain type I, the expectation values of
the tensor multiplets have an interpretation as distances of
five-branes in the extra direction of $M$ theory. Arguments similar to
\horwit, suggest that $M$ theory on $\IR ^1/\IZ _2$ gives, at the origin,
the infinite coupling limit of the type I $SO(32)$ ten-dimensional
string theory.  By ``compactifying'' this theory on the infinite
volume ALE space, we can obtain a finitely coupled type I theory.  In
analogy with the $E_8$ case, we expect that a small $SO(32)$ instanton
turns into an NS five-brane which can wander away from the origin into
$\IR /\IZ _2$, with distance equal to the expectation value of the
tensor multiplet.

Sufficiently far along the Coulomb branch, where the tensor multiplet
has large expectation value, the theory is weakly coupled and we can
apply type I perturbation theory.  However, we can not blow up the ALE
singularity (or turn on a $B$ field) because the corresponding
hypermultiplet of type IIB was projected out.  At the origin of the
Coulomb branch there is the transition point to the Higgs branch,
where the orbifold singularity can be blown up, but we can not
continue to this branch via our perturbative type I orientifold
analysis because the interactions at the origin are too strong.  Via
$S$ duality between the type I and heterotic $SO(32)$ strings in ten
dimensions, there is a heterotic description with coupling
$\lambda_{heterotic}={\V _4\over \lambda_{type I}}$.  For compact $\V
_4$, it is thus possible to describe the Higgs branch, where $\lambda
_{type I}$ is large, in heterotic perturbation theory.  In the
non-compact case, $\lambda _{heterotic}$ blows up.  The conclusion is
that $M$ theory and type I perturbation theory are adequate to
describe the Coulomb branch, while the Higgs branch is described by
the perturbative heterotic string or $F$ theory.

\newsec{Other theories, with five-branes decoupled from the nine-branes}

There are other 6d gauge theories of a type similar to those described
in sect. 5, but with no coupling to the nine-branes: i.e. all $w_\mu
=0$.  The theories are again based on moose diagrams which correspond
to extended Dynkin diagrams and can be constructed for all
$A_{r=2M-1}$, $D_r$, $E_6$, and $E_7$ Dynkin diagrams.  The basic
difference from the theories discussed in sect. 5 can be regarded as a
different action $*'$ of the complex-conjugation operation.  Much as
in \sigg, the gauge group is
\eqn\siggn{\prod _{\mu \in \R '}Sp(v_\mu )\times \prod _{\mu \in
\P '}SO(v_\mu )\times \prod _{\mu \in \C '}U(v_\mu ),}
where we decompose the set of nodes of the extended Dynkin
diagram into subsets as $\{\mu \}=\R ' \oplus \P '\oplus \C '\oplus
\overline {\C '}$, with the subsets in the various cases defined as
follows (with sets not listed empty): For $A_{r=2M-1}$, $\R
'=\{2s|s=0\dots M-1\}$, $\P '=\{2s+1|s=0\dots M-1\}$.  For $D_r$ with
$r$ odd, $\R '=\{0,1, 2s+1|s=1\dots \half (r-3)\}$, $\P '=\{r-1, r,
2s| s=1\dots \half (r-3)\}$.  For $D_r$ with $r$ even, $\R '=\{0,1,
2s+1| s=1\dots \half (r-4)\}$, $\P '=\{2s| s=1\dots \half (r-2)\}$,
$\C '=\{r-1\}$, $\overline{\C '}=\{r\}$.  For $E_6$, $\R
'=\{0,1,3,5\}$, $\P '=\{2,4,6\}$.  For $E_7$, $\R '=\{0,6, 7 \}$, $\P
'=\{ 3\}$, $\C '=\{1,2 \}$, $\overline {\C '}=\{4,5\}$.  The gauge
theory \siggn\ has hypermultiplets corresponding to the links in the
extended Dynkin diagrams: there are hypermultiplets in the $\f{1}{4}
\oplus _{\mu \nu}a_{\mu \nu}(\fund _\mu , \fund _\nu )$, where the
factor in $\f{1}{4}\oplus _{\mu \nu}$ is the same as explained after
\sigg.

The six-dimensional gauge anomaly of the \siggn\ is 
\eqn\ganomn{\cA =\half (\widetilde C_{\mu \nu} V_\nu
+16S '_\mu)\tr F_\mu ^4+\f{3}{2}\widetilde C_{\mu \nu}
\tr F_\mu ^2\tr F_\nu ^2,} where we define $S '_\mu \equiv 1$ for $\mu
\in \R '$, $S '_\mu \equiv -1$ for $\mu
\in \P'$, and $S '_\mu \equiv 0$ for $\mu \in \C '$ or $\overline \C
'$ and $F_{\overline \mu}\equiv F_\mu$ along with $V_{\overline
\mu}\equiv V_\mu \equiv \dim (\fund _\mu )$ and $w_{\overline
\mu}\equiv w_\mu$ for $\overline \mu \in \overline \C$.
The irreducible anomaly in \ganomn\ vanishes provided
\eqn\visn{V_\mu =2Kn_\mu -16G_{\mu \nu}S'_{\nu},}
where $K$ is an {\it arbitrary} integer.  This is similar to the
expressions in sect. 5 except that here there are no hypermultiplets
associated with coupling to nine-branes, all
$w_\mu =0$.  Indeed, introducing $w_\mu$ extra (half) hypermultiplets
in the $\fund _\mu$ would lead to a modification of \ganomn\ analogous
to
\ganomi, but the requirement of canceling the irreducible $\tr F_\mu
^4$ anomalies would then give $w_\mu n_\mu =0$, i.e. all $w_\mu =0$,
because here $16n_\mu S '_\mu =0$ in every case rather than $16n_\mu
S_\mu =32$ as found in sect. 5.  Finally, much as in sect. 5, the
reducible anomaly in \ganomn\ can be cancelled by introducing $r-|\C
'|$ tensor multiplets.  

The simplest example of these theories is the case associated with the
$SU(2)$ extended Dynkin diagram.  The gauge group \siggn\ is
$Sp(K)\times SO(2K+8)$ with two half-hypermultiplets in the $(\fund,
\fund )$, corresponding to the two links in the diagram. $K$ is an
{\it arbitrary} integer. These theories were presented in \Sanom\ as a
class of theories which satisfy the anomaly cancellation conditions
when coupled to gravity.  Here we see that the anomaly can be
cancelled even with gravity decoupled, by including an extra tensor
multiplet.  Another simple example is the case associated with the
$D_4$ extended Dynkin diagram.  The gauge group \siggn\ for $D_4$ is 
$Sp(K)_0\times Sp(K)_1\times SO(4K+16)\times U(2K+8)$ with
half-hypermultiplets in the $({\bf 2K}_i, {\bf 4K+16})$, $i=0,1$, and
a hypermultiplet in the $({\bf 4K+16}, {\bf 2K+8})$, corresponding to
the links.  In addition, there are $r-|\C '|=3$ tensor multiplets. 

Each of the theories \siggn\ with \visn\ and $r-|\C '|$ tensors lead
to an infinite family of 6d RG fixed points, labeled by $K$.  Again,
it is necessary to take the $Sp(v_0)$ factor in \siggn\ to be a global
rather than gauge symmetry to avoid Landau poles on the Coulomb branch.
Also, the $U(1)$ factors are again eliminated.

\subsec{Orientifold construction of the theories}

The $A_1$ case discussed in \Sanom\ and above, with gauge group
$Sp(\half V_0)\times SO(V_1)$, can be constructed via an 
$\Omega \IZ _4$ orientifold.  We take $\alpha =e^{2\pi i/4}$ the
generator of $\IZ _4$ and represent the generator of $\Omega \IZ _4$ by
\eqn\gammix{\gamma _{\Omega \alpha}=\pmatrix{J_{V_0}&0\cr
0&-I_{V_1}},} where, as before, $J$ is the symplectic matrix and $I$
is the identity matrix.  This satisfies the conditions needed for the
case with vector structure so we expect a tensor from the $\IZ _2$
twisted sector.  There are no nine-branes or D5 branes since we do not
have elements $\Omega$ or $\Omega \alpha ^2$.  Because there are no D5
branes or nine-branes, there normally would be no mechanism for
cancelling crosscap terms by branes.  However, if we assume the
existence of ``twisted'' five-branes on which twisted open strings can
end \Julie, we can cancel the crosscap terms.  The tadpoles are found
{\' a} la \GJ\ to be
\eqn\nobtad{{1\over 4\sin ^2{\pi \over 2}}\left(4\sin ^2  {\pi \over
2}(V_0-V_1)+32\right) ^2,}
so $\widetilde C_{\mu \nu}V_\nu =-D_\mu \equiv 16 (-1)^\mu$ for $\mu
=0,1$, i.e. $V_1=V_0+8$ as expected.   

This construction would not work for $A_{2M-1}$ singularities with
$M>1$ because there would be disk terms which could not be cancelled
by crosscaps.  However, the orientifold theories $\Omega
(-1)^{F_L}R_3\times \IZ _{2M}$ studied by \refs{\Julie, \BluZaff},
corresponding to F theory on $(\IC ^2/\IZ _{2M}\times T^2)/\IZ _2$,
can be T-dualized to orientifolds with five-branes and nine-branes.
There is no coupling and hence, no matter from the intersection of the
two kinds of branes.  Such a construction can yield the $A_{2M-1}$
case of
\visn\ with $K=0$, which gives $V_{\mu \in \R '}=0$ and $V_{\mu \in
\P'}=8$, i.e. $SO(8)^{\otimes M}$, and the required $2M-1$ extra
tensors.  Indeed, this spectrum for the $M=1,2,3$ cases can easily be
read off the results of \Julie, adjusted to the non-compact case. The
solutions of
\Julie\ were not proven to be the most general solutions, and it would
be interesting to see if all of the above models without coupling to
nine-branes could be obtained from this construction.  It would, of
course, be necessary to include twisted open strings to cancel the
tadpoles.  Because of the lack of coupling to nine-branes, the
five-branes of these theories cannot be interpreted as small instantons
and any heterotic dual description would have to be highly unusual.

\newsec{Building other 6d moose models}

We will make some general observations about how to build 6d theories
for which the irreducible anomaly can be cancelled.  Consider a theory
with a product gauge group, with some matter fields coupling the
representations.  We focus on two factors out of the possibly much
larger theory, groups $G_1\times G_2$, with some matter in
representations $(r_1, r_2)$ coupling them.

Consider first the case of $U(v_1)\times U(v_2)$ coupled by $n_c$
hypermultiplets in the $(\fund ,
\fund )$.  There could also be additional matter multiplets in either
gauge group.  In order to cancel the irreducible $\tr F_i^4$
anomalies, it is necessary to have $2v_1-n_cv_2-P_1=0$ and
$2v_2-n_cv_1-P_2=0$, where $P_1\geq 0$ are contributions associated
with matter charged only under $U(v_i)$.  In order to have $v_i\geq
0$, it is necessary to have $n_c=1$.  The $v_i$ then satisfy $2v_1\geq
v_2\geq \half v_1$.  Also, $P_1< 2v_1$ and $P_2< 2v_2$; thus the
$U(v_i)$ can generally have matter only in the fundamental and/or two
index symmetric or antisymmetric tensor representations.  Further, it
will generally not be possible to have other types of mixed
representations.  For example, if there were a matter field in the
$(\fund , \asym)$ rather than in the $(\fund,
\fund)$, it would be necessary to have $2v_1-\half v_2(v_2-1)=P_1\geq
0$ and $2v_2-v_1(v_2-8)=P_2\geq 0$, which does not have a solution for
$v_2>8$.

Consider next the case of $SO(v_1)\times SO(v_2)$ coupled with $n_c$
multiplets in the $(\fund, \fund)$ and possibly additional matter
multiplets in either group.  In order to cancel the irreducible $\tr
F_i^4$ anomalies, it is necessary to have $v_1-8-n_cv_2-P_1=0$ and
$v_2-8-n_cv_1-P_2=0$.  For any $n_c\neq 0$, there is no solution of
these equations with $v_i\geq 0$ if the $P_i\geq 0$, suggesting that
$SO$ groups can never be coupled directly to other $SO$ groups. 
However there is a loophole here:  spinors of $SO$ groups can lead to a
{\it negative} contribution for the $P_i$ \ref\erler{J. Erler,
hep-th/9304104, J. Math. Phys. 35 (1994) 1819.}.

Consider next $Sp(v_1)\times Sp(v_2)$ coupled with $n_c$ multiplets in
the $(\fund , \fund )$ and possibly additional matter in either
group. In order to cancel the irreducible anomalies, it is necessary
to have $2v_1+8-2n_cv_2=P_1\geq 0$ and $2v_2+8-2n_cv_2=P_2\geq 0$.
For $n_c=1$, this requires $P_1+P_2=16$; this case was discussed in
\obrane -- it arises for small instantons at a $\IZ _2$ orbifold
singularity with possible vector structure.  There can also be
solutions with $n_c>1$ provided $P_i<8$.

For $U(v_1)\times SO(v_2)$ coupled with $n_c$ matter fields in the
$(\fund , \fund)$ and possibly additional matter, the irreducible
anomalies cancel provided $2v_1-n_cv_2=P_1\geq 0$ and
$v_2-8-n_cv_1=P_2\geq 0$.  It is thus necessary to have $n_c\leq 1$.
For $U(v_1)\times Sp(v_2)$ coupled with $n_c$ matter fields in the
$(\fund , \fund)$ and possibly additional matter, the irreducible
anomalies cancel provided $2v_1-2n_cv_2=P_1\geq 0$ and
$v_2+8-n_cv_1=P_2\geq 0$, giving $n_c\leq 1$ and $P_2\geq 8$.  For
$SO(v_1)\times Sp(v_2)$ coupled with $n_c$ half-hypermultiplets in the
$(\fund , \fund)$ and possibly additional matter, the irreducible
anomalies cancel provided $v_1-8-n_cv_2=P_1\geq 0$ and $2v_2+8-\half
n_cv_1=P_2\geq 0$, giving $n_c\leq 2$ and $P_2\geq 8$.

\appendix{A}{Some $\Gamma _G$ representation theory}

We briefly discuss the representations of $\Gamma _G$ for $G=D_r$ and
$G=E_6$ to illustrate the decompositions $\{\mu \}=\R \oplus
\P\oplus \C \oplus \overline {\C}$ and the conjugacy classes
(i.e. that there are $r-2|\C |$ classes which contain their own
inverses and $|\C|$ pairs of classes which do not).

The $\Gamma _{G=D_r}$ dihedral group is generated by elements $\alpha$
and $\beta$ with $\alpha ^2=\beta ^{r-2}=C$ and $\beta \alpha \beta
=\alpha$, where $C$ is the central element $-1$ of $SU(2)$.  
The four representations with $n_\mu =1$ are given by $\alpha _\mu
=(-1)^\mu $, $\beta _\mu =1$ for $\mu =0,1$ and $\alpha _\mu =(-1)^\mu
e^{i\pi r/2}$, $\beta _\mu =-1$ for $\mu =r-1,r$.  The representations
with $n_\mu =2$ are given by 
\eqn\tdreps{\alpha _\mu =e^{i\pi (\mu -1)/2}\pmatrix{1&0\cr 0&-1}, \qquad 
\beta _\mu =\pmatrix{\cos \theta _\mu &\sin \theta _\mu \cr
-\sin \theta _\mu &\cos \theta _\mu }, \quad \theta _\mu \equiv {\pi
(\mu -1)\over r-2},} $\mu =2, \dots r-2$.  For $D_r$ with $r$ even,
the conjugacy classes are: $[1]$, $[-1]$, $[\beta ^k,
\beta ^{-k}]$, $0\leq k\leq r-2$, $[\alpha ]$, and $[\alpha \beta ]$,
which are all self-conjugate (contain their inverse).  For $r$ odd,
the classes are the same but $[\alpha ]$ and $[\alpha \beta ]$ are now
conjugate to each other (contain the other's inverses).  

$\Gamma _{E_6}$ is generated by elements $\alpha$, $\beta$, and $\rho$
with $\alpha ^2=\beta ^2=\rho ^3=C$, $\alpha \rho =\rho \beta \alpha$,
$\beta \rho =\rho \alpha$, $\beta \alpha \rho=\rho \beta$, and $\beta
\alpha \beta =\alpha$.  The
representations $\mu =0, 1, 5$ with $n_\mu =1$ have $\alpha _\mu
=\beta _\mu =1$ and $\rho _\mu =1$, $\omega$, $\omega ^2$,
respectively, where $\omega \equiv e^{2\pi i /3}$.  The representation
$\mu =2$ has $\alpha _\mu =i\sigma _1$, $\beta _\mu =i\sigma _3$, and
$\rho _\mu ={e^{2\pi i /8}\over \sqrt{2}}\pmatrix{-i&-i\cr -1 &1}$ and
is pseudo-real.  The other two representations with $n_\mu =2$, $\mu
=4$ and $\mu =6$, differ from $\mu =2$ in that $\rho _4=\omega
\rho _2$ and $\rho _6=\omega ^2 \rho _2$; these two representations
are complex conjugates of each other.  Finally, the representation
$\mu =3$ is given by
\eqn\evirepiii{\alpha _3=\pmatrix {-1 &0 & 0\cr 0& -1 &0 \cr 0& 0& 1}\quad
\beta _3=\pmatrix {-1 &0 &0 \cr 0& 1 &0 \cr 0&0& -1}\quad \rho
_3=\pmatrix{0&1&0\cr 0&0&1\cr 1&0&0}.}
The conjugacy classes of $\Gamma _{E_6}$ are $[1]$, $[-1]$, $[\rho ]$,
$[\rho ^2 ]$, $[\alpha ]$, $[-\rho ]$, and $[-\rho ^2 ]$, with $[\rho
]$ and $[\rho ^2 ]$ conjugate and $[-\rho ]$ and $[-\rho ^2 ]$
conjugate.

\centerline{{\bf Acknowledgments}}

The work of J.B. is supported in part by NSF PHY-9513835. The work of
K.I. is supported by NSF PHY-9513835, the W.M. Keck Foundation, an
Alfred Sloan Foundation Fellowship, and the generosity of Martin and
Helen Chooljian.

\listrefs
\end